\documentclass[conference]{IEEEtran}
\usepackage{amsfonts,listings,courier,pgfplotstable,times,xspace,tikz,amssymb,
  amsmath,comment,stmaryrd,subcaption,multirow,wrapfig,pifont,bigdelim}
\usetikzlibrary{shapes,arrows,chains,fit,backgrounds}

\tikzstyle{bef}=[align=left,font=\ttfamily,node distance=3mm,text width=\textwidth*.4]
\tikzstyle{aft}=[align=left,font=\ttfamily,node distance=3mm,text width=\textwidth*.5]
\tikzstyle{acsl}=[inner sep=0mm,fill=blue!15]
\tikzstyle{genc}=[inner sep=0mm,fill=blue!15]

\tikzstyle{common}=[node distance=15mm,align=center]
\tikzstyle{test}=[draw,trapezium,trapezium left angle=70,
  trapezium right angle=-70,common]
\tikzstyle{op}=[draw,common]
\tikzstyle{data}=[draw,common,ellipse,node distance=35mm]
\tikzstyle{ce}=[data]
\tikzstyle{arrow}=[draw,->]
\tikzstyle{darrow}=[draw,o->]

\newtheorem{definition}{Definition}

\newcommand*\circled[1]{\tikz[baseline=(char.base)]{
    \node[shape=circle,draw,inner sep=.5mm,font=\bfseries] (char) {#1};}}

\newcommand{\ce}[1]{{\scriptsize{$\langle$}}#1{\scriptsize{$\rangle$}}}
\newcommand{\fassert}{\lstinline'fassert'\xspace}
\newcommand{\fassume}{\lstinline'fassume'\xspace}
\newcommand{\cmark}{\ding{51}}
\newcommand{\xmark}{\ding{55}}
\newcommand{\ok}{\cmark\xspace}
\newcommand{\ko}{\xmark\xspace}

\newcommand{\modif}[1]{{#1}\xspace}

\newcommand{\stady}{\textsc{StaDy}\xspace}
\newcommand{\framac}{\textsc{Frama-C}\xspace}
\newcommand{\pathcrawler}{\textsc{Path\-Craw\-ler}\xspace}
\newcommand{\acsl}{\textsc{acsl}\xspace}
\newcommand{\eacsl}{\textsc{e-acsl}\xspace}
\newcommand{\Wp}{\textsc{Wp}\xspace}
\newcommand{\Dafny}{\textsc{Dafny}\xspace}
\newcommand{\Value}{\textsc{Value}\xspace}
\newcommand{\sante}{\textsc{sante}\xspace}
\newcommand{\eacsltoc}{\textsc{e-acsl2c}\xspace}

\newcommand{\NC}{\ensuremath{\mathrm{NC}}\xspace}
\newcommand{\NCD}{\ensuremath{\mathrm{NCD}}\xspace}
\newcommand{\GSW}{\ensuremath{\mathrm{GSW}}\xspace}
\newcommand{\SSW}{\ensuremath{\mathrm{SSW}}\xspace}  
\newcommand{\CWD}{\ensuremath{\mathrm{SWD}}\xspace}  
\newcommand{\SWD}{\ensuremath{\mathrm{SWD}}\xspace}  
\newcommand{\GSWD}{\ensuremath{\mathrm{GSWD}}\xspace}  
\newcommand{\SSWD}{\ensuremath{\mathrm{SSWD}}\xspace}  
\newcommand{\NCCE}{\ensuremath{\mathrm{NCCE}}\xspace}
\newcommand{\CWCE}{\ensuremath{\mathrm{SWCE}}\xspace}  
\newcommand{\SWCE}{\ensuremath{\mathrm{SWCE}}\xspace}  
\newcommand{\GSWCE}{\ensuremath{\mathrm{GSWCE}}\xspace}
\newcommand{\SSWCE}{\ensuremath{\mathrm{SSWCE}}\xspace}

\newcommand{\nc}{\textsf{nc}\xspace}
\newcommand{\cw}{\textsf{sw}\xspace}  
\newcommand{\no}{\textsf{no}\xspace}

\newcommand{\citeframac}{\cite{Frama-C}\xspace}
\newcommand{\citepathcrawler}{\cite{PathCrawler}\xspace}

\newcommand{\citesante}{\cite{DBLP:conf/sac/ChebaroKGJ12}\xspace}

\newcommand{\C}{{\ensuremath\mathcal{C}\xspace}}
\newcommand{\A}{{\ensuremath\mathcal{A}\xspace}}
\lstdefinelanguage{pretty-ACSL}{%
  escapechar={},
  literate=
   {==}{{$==$}}2
   {==>}{{$\Rightarrow$}}1
   {integer\ i}{{i$\,\in \mathbb{Z}\,$}}4
   {integer\ j}{{j$\,\in \mathbb{Z}\,$}}4
   {integer\ k}{{k$\,\in \mathbb{Z}\,$}}4
   {integer\ m}{{m$\,\in \mathbb{Z}\,$}}4
   {integer\ l}{{l$\,\in \mathbb{Z}\,$}}4
   {\\forall}{{$\forall$}}1
   {\\exists}{{$\exists$}}1
   {integer}{{$\mathbb{Z}$}}1
   {real}{{$\mathbb{R}$}}1
   {&&}{{$\wedge$}}1
   {||}{{$\vee$}}1
   {!=}{{$\neq$}}1
   {<}{{$<$}}1
   {<=}{{$\le~$}}1
   {>}{{$>$}}1
   {>=}{{$\ge~$}}1
   {<==>}{{$\Leftrightarrow$}}1,
  morekeywords={assert,assigns,assumes,axiom,axiomatic,behavior,behaviors,
    boolean,breaks,complete,continues,data,decreases,disjoint,ensures,
    exit_behavior,ghost,global,inductive,invariant,lemma,logic,loop,
    model,predicate,reads,requires,sizeof,strong,struct,terminates,
    type,union,variant,uchar,byte,typically,\\result,\\old,\\at,\\valid,
    \\separated,\\nothing,Pre,\\sum,\\numof},
  alsoletter={\\},
  morecomment=[l]{//}
}
\lstnewenvironment{pretty-codeACSL}{\lstset{language=pretty-ACSL,stepnumber=0}}{\smallskip}

\lstdefinelanguage{ACSL}{%
  escapechar={},
  literate=,
  morekeywords={assert,assigns,assumes,axiom,axiomatic,behavior,behaviors,
    boolean,breaks,complete,continues,data,decreases,disjoint,ensures,
    exit_behavior,ghost,global,inductive,invariant,lemma,logic,loop,
    model,predicate,reads,requires,sizeof,strong,struct,terminates,
    type,union,variant,uchar,byte,typically,\\result,\\old,\\at,\\valid,
    \\separated,\\nothing,Pre,\\exists,\\forall,\\sum,\\numof},
  alsoletter={\\},
  morecomment=[l]{//}
}
\lstnewenvironment{codeACSL}{\lstset{language=ACSL,stepnumber=0}}{\smallskip}

\lstdefinestyle{pretty-c}{language={[ANSI]C},%
  alsolanguage=pretty-ACSL,%
  moredelim={*[l]{//}},%
  deletecomment={[s]{/*}{*/}},
  moredelim={*[l]{//@}},%
}

\lstdefinestyle{c}{language={[ANSI]C},%
  alsolanguage=ACSL,%
  moredelim={*[l]{//}},%
  deletecomment={[s]{/*}{*/}},
  moredelim={*[l]{//@}},%
}

\IEEEoverridecommandlockouts
\begin{document}



\title{Your Proof Fails? Testing Helps to Find the Reason
}
\author{
  \IEEEauthorblockN{
    Guillaume Petiot\IEEEauthorrefmark{1}\IEEEauthorrefmark{2},
    Nikolai Kosmatov\IEEEauthorrefmark{1},
    Bernard Botella\IEEEauthorrefmark{1},
    Alain Giorgetti\IEEEauthorrefmark{2}\IEEEauthorrefmark{3} and
    Jacques Julliand\IEEEauthorrefmark{2} 
  }
  \IEEEauthorblockA{
    \IEEEauthorrefmark{1}
    CEA, LIST, Software Reliability Laboratory,
    PC 174, 91191 Gif-sur-Yvette France\\
    Email: \texttt{firstname.lastname@cea.fr}
  }
  \IEEEauthorblockA{
    \IEEEauthorrefmark{2}
    FEMTO-ST/DISC, University of Franche-Comt\'e,
    25030 Besan\c{c}on Cedex France\\
    Email: \texttt{firstname.lastname@femto-st.fr}
  }
  \IEEEauthorblockA{
    \IEEEauthorrefmark{3}
    INRIA Nancy - Grand Est, CASSIS project, 54600 Villers-l\`es-Nancy France
  }
}

\date{\today}
\maketitle

\begin{abstract}
  Applying deductive verification to formally prove that a program respects its
  formal specification is a very complex
  and time-consuming task due in particular to the lack of feedback in case of
  proof failures.
  Along with a non-compliance between the code and its specification (due to an error 
  in at least one of them),
  possible reasons of a proof failure include a missing or too weak \modif{specification} for a called function or a loop,
  and lack of time or simply incapacity of the prover to finish a particular
  proof.
  This work proposes a new methodology where test generation helps
  to identify the reason of a proof failure and to exhibit 
  a counter-example clearly illustrating the issue.
  We describe how to transform an annotated C program 
  into C code suitable for testing and illustrate the benefits of the 
  method on comprehensive examples.
  The method has been implemented in \stady, a plugin of the software analysis platform
  \framac.
  Initial experiments show that detecting non-compliances and contract weaknesses 
  allows to precisely diagnose most proof failures.
\medskip

 \textbf{Keywords:} deductive verification, test generation, 
 specification, proof failure, non-compliance detection,
 contract weakness detection, Frama-C
\end{abstract}

\vspace{-2mm}
\section{Introduction}
\label{sec:intro}
\vspace{-1mm}

Among
formal verification techniques, \emph{deductive verification}
consists in establishing a rigorous mathematical proof that a given
program meets its specification. When no confusion is possible, one also says
that deductive verification consists in ``proving a program''. It requires that
the program comes with a formal specification, usually given in special comments 
called \emph{annotations,} including function contracts (with
pre- and postconditions) and loop contracts (with loop variants and invariants).
The \emph{weakest precondition calculus} proposed by
Dijkstra~\cite{DBLP:books/ph/Dijkstra76} reduces any deductive verification
problem to 
establishing the validity of first-order
formulas called \emph{verification conditions}.

In modular deductive verification of a function $f$ calling another function
$g$, the roles of the pre- and postconditions of $f$ and of the callee $g$ are
dual.
The precondition of $f$ is assumed and its  postcondition must be proved, while
at  any call of $g$ in $f$, 
the precondition of
$g$ must be proved before the call and its postcondition is assumed after the
call.
\modif{ The situation for a function $f$ with one call to $g$ is presented in
Fig.~\ref{fig:verif-func-call}.
An arrow in this figure informally indicates that its initial point provides a
hypothesis for a proof of its final point.
For instance, the precondition $\textit{Pre}_f$ of $f$ and the postcondition
$\textit{Post}_g$ of $g$ provide hypotheses for a proof of the postcondition
$\textit{Post}_f$ of $f$.
The called function $g$ is proved separately.
The verification of the loop invariant $I$ of a loop in $f$ is illustrated by
Fig.~\ref{fig:loop}: $I$ must be proved to hold initially before the first
loop iteration, and $I\wedge\neg b$ is assumed 
after exiting the loop.
In addition, the preservation of the loop invariant $I$ by each unique iteration
of the loop must be established during the proof of $f$.
(Loop termination, not illustrated in Fig.~\ref{fig:loop}, can be proved
as well.)}

\begin{wrapfigure}[29]{r}{46mm}
 \begin{subfigure}{46mm}
%
%
\begin{footnotesize}
\begin{tikzpicture}[scale=0.7]
\node[text width=3cm] (fPre) at (0,4){\lstinline{//} 
 $\textit{Pre}_f$ \lstinline{assumed}}; 
\node[text width=3cm] (fBegin) at (0,3.5){\lstinline{f(<args>){}};
\node[text width=3cm] (Code1) at (0,3){\lstinline{  code1;}};
\node[text width=3cm] (gPre) at (0,2.5){\lstinline{//} $
 \textit{Pre}_g$ \lstinline{to be proved}}; 
\node[text width=3cm] (gCall) at (0,2){\lstinline{  g(<args>);}};
\node[text width=3cm] (gPost) at (0,1.5){\lstinline{//} 
 $\textit{Post}_g$ \lstinline{assumed}}; 
\node[text width=3cm] (Code2) at (0,1){\lstinline{  code2;}}; 
\node[text width=3cm] (fEnd) at (0,0.5){\lstinline'}'};
\node[text width=3cm] (fPost) at (0,0){\lstinline{//} 
 $\textit{Post}_f$ \lstinline{to be proved}};
\draw[->,>=latex,thick,color=black] (fPre) to[bend right=86] (gPre);
\draw[->,>=latex,thick,color=black] (gPost) to[bend right=86] (fPost);
\draw[->,>=latex,thick,color=black] (fPre) to[bend right=86] (fPost);
\end{tikzpicture}
\caption{called function $g$} 
\label{fig:verif-func-call}
\end{footnotesize}
\end{subfigure}
 \begin{subfigure}{46mm}
\begin{footnotesize}
\begin{tikzpicture}[scale=0.7]
\node[text width=3cm] (fPre) at (0,6){\lstinline{//}
 $\textit{Pre}_f$ \lstinline{assumed}};
\node[text width=3cm] (fBegin) at (0,5.5){\lstinline{f(<args>){}};
\node[text width=3cm] (Code1) at (0,5){\lstinline{  code1;}};
\node[text width=3cm] (lPre) at (0,4.5){\lstinline{//}
 $I$ \lstinline{to be proved}};
\node[text width=3cm] (lBegin) at (0,4){
 \lstinline'  while('$b$\lstinline'){'}; 
\node[text width=3cm] (iPre) at (0,3.5){\lstinline{//}
 $I\, \wedge\, b$\lstinline{ assumed}};
\node[text width=3cm] (Code3) at (0,3){\lstinline'    code3;'};
\node[text width=3cm] (iPost) at (0,2.5){\lstinline{//}
 $I$ \lstinline{to be proved}};
\node[text width=3cm] (lEnd) at (0,2){\lstinline'  }'};
\node[text width=3cm] (lPost) at (0,1.5){\lstinline{//}
 $I\, \wedge\,\neg b$\lstinline{ assumed}};
\node[text width=3cm] (Code2) at (0,1){\lstinline{  code2;}};
\node[text width=3cm] (fEnd) at (0,0.5){\lstinline'}'};
\node[text width=3cm] (fPost) at (0,0){\lstinline{//}
 $\textit{Post}_f$ \lstinline{to be proved}};
\draw[->,>=latex,thick,color=black] (fPre) to[bend right=90] (lPre);
\draw[->,>=latex,thick,color=black] (lPost) to[bend right=90] (fPost);
\draw[->,>=latex,thick,color=black] (iPre) to[bend right=90] (iPost);
\draw[->,>=latex,thick,color=black] (fPre) to[bend right=90] (iPost);
\draw[->,>=latex,thick,color=black] (fPre) to[bend right=90] (fPost);
\end{tikzpicture}
\vspace{-2mm}
\caption{loop} 
\label{fig:loop}
\end{footnotesize}
\end{subfigure}
\caption{Verification of a function $f$ with a callee or a loop}
\end{wrapfigure}

To reflect the fact that some contracts become hypotheses
during deductive verification of $f$
we use the term \emph{subcontracts for $f$} 
to designate contracts of called functions and loops in $f$.

\textbf{Motivation.} 
One of the most important difficulties in deductive verification is the manual
processing of proof failures 
by the verification engineer since proof failures may have several
causes. 
Indeed, a failure to prove $\textit{Pre}_g$ in Fig.~\ref{fig:verif-func-call} may be due to
a \emph{non-compliance} of the code to the specification:
an error in the code \lstinline'code1', or a 
wrong specification $\textit{Pre}_f$ or $\textit{Pre}_g$ itself
that may incorrectly formalize the requirements.
The verification can also remain inconclusive because of 
a \emph{prover incapacity} to finish
a particular proof within an allocated time. 
%
%
In many cases, it is extremely difficult for the verification engineer 
to decide how to proceed: 
either suspect a non-compliance and look for an error in the code or check the specification, 
or suspect a prover incapacity, give up automatic proof and try to achieve an
interactive proof with a proof assistant (like \textsc{Coq} \cite{coq}).

A failure to prove the postcondition $\textit{Post}_f$ 
(cf. Fig.~\ref{fig:verif-func-call}) is even more complex to analyze: along with a
prover incapacity or a non-compliance due to errors in the pieces of code
\lstinline'code1' and \lstinline'code2' or an incorrect
specification $\textit{Pre}_f$ \modif{or $\textit{Post}_f$,} the failure can
\modif{also} result from a too weak postcondition $\textit{Post}_g$ of $g$, that
does not fully express the intended behavior of $g$. Notice that in this last case, the proof of $g$ can
still be successful. 
The current automated tools for program proving
do not provide a precise 
indication on the reason of the proof failure.
The most advanced tools (like \Dafny \cite{Leino/FIDE14}) produce a
counter-example extracted from the underlying solver
without saying directly if the verification engineer should look for 
a non-compliance,
or strengthen subcontracts (and which one of them), 
or consider adding additional lemmas or using interactive proof.
So the verification engineer must basically consider all possible reasons one
after another, maybe also trying a very costly interactive proof.
For a loop, 
the situation is similar
and offers an additional challenge:
to prove the invariant preservation, whose failure 
can be due to several reasons as well.
%
%

The motivation of this work is twofold. First, we want to provide the
verification engineer with a more precise feedback indicating the reason of each
proof failure. Second, we look for a counter-example that either confirms the
non-compliance and demonstrates that the unproven predicate can indeed fail on a
test datum, or confirms a subcontract weakness
showing on a test datum which subcontract is insufficient.

\textbf{Approach and goals.}
We propose to
use advanced test generation techniques in order to diagnose 
a proof failure and produce counter-examples.
Their usage requires a translation of the annotated
C program
into an executable C code suitable for testing.
Previous works
addressed the generation of counter-examples only for
non-compliance~\cite{Petiot/TAP14} and proposed a rule-based formalization of
annotation translation
in that case~\cite{Petiot/SCAM14}.
The cases of subcontract weakness remained undetected and indistinguishable from
a prover incapacity.
The overall goal of the present work is to provide a methodology for a more
precise  identification of proof failure reasons in all these cases, to
implement it and to evaluate it in practice.
The proposed method is composed of two steps. The first step looks for
non-compliance. If no non-compliance is detected, the second step looks for a
subcontract weakness.
Another goal is to make \modif{this method} automatic and suitable for a non-expert verification
engineer.
\modif{
Following the modular verification approach,
we assume that the called functions respect their contracts.
To simplify the presentation, we also assume that 
the loops preserve their loop invariants, and focus
on other proof failures occurring during modular verification of $f$. 
(The proposed detection techniques can be adapted to the verification of a loop contract.)
}

\textbf{The contributions} 
of this paper include:
\vspace{-2mm}
\begin{itemize}
\item a classification of proof failures into three categories: non-compliance,
subcontract weakness and prover incapacity,
\item a definition of 
counter-examples for the first two categories, 
\item a new program transformation technique for the diagnosis 
of a subcontract weakness by testing 
(in addition to the one previously proposed for
non-compliance~\cite{Petiot/SCAM14}), 
\item \modif{a complete testing-based methodology for diagnosis of proof failures and
generation of counter-examples, 
suggesting possible actions for each category,
illustrated on several comprehensive examples,}
\item an implementation of the proposed solution in a tool called \stady, and
\item experiments showing its capacity of diagnosis of proof failures.
\end{itemize}
\vspace{-2mm}

\textbf{Paper outline.} 
Sections~\ref{sec:framac} and~\ref{subsec:example}  respectively present the
tools used in this work and an illustrative example.
Section~\ref{sec:definitions} defines the categories of proof failures and
counter-examples, and presents program transformations for their identification.
The complete methodology for the diagnosis of proof failures is presented in
Section~\ref{sec:global-method}. 
Our implementation and experiments are described in
Sec.~\ref{sec:implementation}.
Finally, Sections~\ref{sec:related} and~\ref{sec:conclusion}
 present some related works and a conclusion.

\vspace{-2mm}
\section{\framac Toolset}
\label{sec:framac}
\vspace{-2mm}

This work is realized in the context of the \framac toolset.
\framac~\citeframac is a platform dedicated to analysis of C
programs that includes various source code analyzers in separate plugins.
The \Value plugin performs value analysis by abstract interpretation.
The \Wp plugin performs weakest precondition calculus for deductive
verification of C programs.
Several automatic SMT solvers can be used to prove the 
verification conditions generated by \Wp. In this work  we use
\textsc{Alt-Ergo} 0.99.1 and \textsc{CVC3} 2.4.1.
\framac also includes plugins for control-flow and program dependency graph construction,
program slicing, impact analysis, test generation, etc.

To express properties over C programs, 
\framac offers a behavioral specification language named
\acsl~\cite{ACSL,Frama-C}.
\acsl annotations play a central role in communication between plugins:
any analyzer can both add annotations to be verified by other ones and notify 
other plugins about its own analysis results by changing an annotation status.
The status can indicate that the annotation is valid, valid under conditions,
invalid or undetermined, and which analyzer established that result.

For combinations with dynamic analysis, \framac also supports  \eacsl
\cite{Delahaye/SAC13,E-ACSL}, a rich executable subset of \acsl suitable for
\emph{runtime assertion checking}.
\eacsl can express function
contracts (pre/postconditions, guarded behaviors, completeness and disjointness
of behaviors), assertions and loop contracts (variants and invariants). It
supports quantifications over bounded intervals of integers, mathematical
integers and memory-related constructs (e.g. on validity and initialization).
It comes with an instrumentation-based translating plugin, called
\eacsltoc, that translates annotations into additional C code in order to
evaluate annotations at runtime and report failures.
\modif{
Important differences  between a translation for runtime assertion
checking and a translation for test generation
(e.g. to support unbounded integer arithmetics in \eacsl and
some specific annotations) \cite{Petiot/SCAM14} make \eacsltoc inadequate for our work and create the
need for a dedicated translation tool.}

For test generation, this work relies on \pathcrawler~\cite{PathCrawler},  
a Dynamic Symbolic Execution 
testing tool, combining \textit{conc}rete and symb\textit{olic} execution.
\pathcrawler is based on a specific constraint solver,
\textsc{Colibri}, that implements
advanced features such as floating-point and modular integer arithmetics support. 
\pathcrawler provides coverage strategies like
{\em k-path} (feasible paths with at most $k$ consecutive loop iterations)
and {\em all-paths} (all feasible paths without any limitation on loop
iterations). \pathcrawler is {\em sound}, meaning that each test case
activates the test objective for which it was
generated. This is verified by concrete execution. 
\pathcrawler is also {\em complete} in the following sense: 
when the tool manages to explore all feasible paths of the program, 
all features of the program are supported by the tool and  constraint solving
terminates for all paths, the
absence of a test for some test objective means that the test objective is
infeasible, since the tool does not approximate path constraints \cite[Sec.\,3.1]{PathCrawler}.

\vspace{-2mm}
\section{Illustrating Example}
\label{subsec:example}
\vspace{-2mm}
We illustrate the issues arising in deductive verification of programs and the
solutions we propose on the example of C program of
Fig.~\ref{fig:rgf1}. It comes from an ongoing work on formal specification and
deductive verification%
~\cite{ggp15}
and implements an 
algorithm proposed in \cite[page 235]{DBLP:books/x/ftx2010}.
The example of Fig.~\ref{fig:rgf1} 
concerns the generation of 
\emph{Restricted Growth Functions} (RGF),
defined by the property expressed by the \acsl
predicate \lstinline{is_rgf} on lines 1--2 of Fig.~\ref{fig:rgf1}, where the RGF
$a$ is represented by the C array 
of its values. 
For convenience of the reader, 
some \acsl notations are  replaced by mathematical symbols
(e.g. keywords \lstinline[style=c]{\exists}, \lstinline[style=c]{\forall} and
\lstinline[style=c]{integer} are respectively denoted by $\exists$, $\forall$ and $\mathbb{Z}$).


\begin{figure}[tb]
  \centering
\begin{lstlisting}
/*@ predicate is_rgf(int *a, integer n) =
  a[0] == 0 && \forall integer i; 1 <= i < n ==> (0 <= a[i] <= a[i-1]+1); */

/*@ lemma max_rgf: \forall int* a; \forall integer n;
  is_rgf(a, n) ==> (\forall integer i; 0 <= i < n ==> a[i] <= i); */

/*@ requires n > 0;
    requires \valid(a+(0..n-1));
    requires 1 <= i <= n-1;
    requires is_rgf(a,i+1);
    assigns a[i+1..n-1];
    ensures is_rgf(a,n); */
void g(int a[], int n, int i) {
 int k;
 /*@ loop invariant i+1 <= k <= n;
     loop invariant is_rgf(a,k);
     loop assigns k, a[i+1..n-1];
     loop variant n-k; */
 for (k = i+1; k < n; k++) a[k] = 0;
}

/*@ requires n > 0;
    requires \valid(a+(0..n-1));
    requires is_rgf(a,n);
    assigns a[1..n-1];
    ensures is_rgf(a,n);
    ensures \result == 1 ==>
     \exists integer j; 0 <= j < n &&
      (\at(a[j],Pre) < a[j] &&
       \forall integer k; 0 <= k < j ==> \at(a[k],Pre) == a[k]); */
int f(int a[], int n) {
 int i,k;
 /*@ loop invariant 0 <= i <= n-1;
     loop assigns i;
     loop variant i; */
 for (i = n-1; i >= 1; i--)
  if (a[i] <= a[i-1]) { break; }
 if (i == 0) { return 0; } // Last RGF.
 //@ assert a[i]+1 <= 2147483647;
 a[i] = a[i] + 1;
 g(a,n,i);
 /*@ assert \forall integer l; 0 <= l < i ==> \at(a[l],Pre) == a[l]; */
 return 1;
}
\end{lstlisting}
  \caption{Successor function for restricted growth functions (RGF)\label{fig:rgf1}}
  \vspace{-5mm}
\end{figure}

Fig.~\ref{fig:rgf1} shows a main function \lstinline{f} 
and an auxiliary function \lstinline{g}.
The precondition of \lstinline{f} states that \lstinline{a} is a valid array 
of size  \lstinline{n>0} (lines 22--23) and must be an RGF (line 24).
The postcondition states 
that the function is only allowed to modify the values of array \lstinline{a} except
the first one \lstinline{a[0]} (line 25), 
and that the generated array \lstinline{a} is still an RGF (line 26).
Moreover, if the function returns 1 then the generated RGF \lstinline{a} must
respect an additional property
(lines 27--30).
Here \lstinline{\at(a[j],Pre)} denotes the value of \lstinline{a[j]} in the 
\lstinline{Pre} state, i.e. before the function starts execution.




We focus now on the body of the function \lstinline{f} in
Fig.~\ref{fig:rgf1}. 
The loop on lines 36--37 goes through the array from right
to left to find the rightmost non-increasing element, that is,
the maximal  array index \lstinline{i} such that \lstinline{a[i] <= a[i-1]}. 
If such an index \lstinline{i} is found, 
the function increments \lstinline{a[i]} (line 40) and fills out the rest
of the array with \lstinline{0}'s (call to \lstinline{g}, line 41).
The loop contract (lines 33--35) specifies the interval of values of the loop variable,
the variable that the loop can modify as well as a loop variant 
that can be used to ensure the termination of the loop.
The loop variant expression must be non-negative whenever an iteration starts,
and strictly decrease after it.

The function \lstinline{g} is used to fill the array with zeros to the right of index \lstinline{i}.
In addition to size and validity constraints (lines 7--8),
its precondition requires that the elements of \lstinline{a} up to index  \lstinline{i}
form an RGF (lines 9--10). 
The function is allowed to modify the elements of \lstinline{a} starting from the index \lstinline{i+1} (line 11)
and generates an RGF (line 12).
The loop invariants 
indicate the value interval of the loop variable \lstinline{k} (line 15), 
and state that the property \lstinline{is_rgf} is satisfied
up to \lstinline{k} (line 16).
This invariant allows a deductive verification tool to deduce the postcondition.
The annotation \lstinline{loop assigns} (line 17) says that the only values the loop can
change are \lstinline{k} and the elements of \lstinline{a} starting from the index \lstinline{i+1}.
The term \lstinline{n-k} is a variant of the loop (line 18).

The \acsl lemma \lstinline'max_rgf' on lines 4--5 
states
that if an array is an RGF, then each of its elements is at most equal to its
index.
This lemma is not proved as such by \Wp but can be used to ensure the absence of
overflow at line 40.

\modif{
The functions of Fig.~\ref{fig:rgf1} can be fully proved using \Wp. 
Suppose now this example contains one of the  following four mistakes: 
the verification engineer \emph{either} forgets the precondition on line 24, 
\emph{or} writes the wrong assignment \lstinline[style=c]'a[i]=a[i]+2;' on line 40,
\emph{or} puts a too general clause \lstinline[style=c]'loop assigns i,a[1..n-1];' on line 34,
\emph{or} forgets to provide the lemma  on lines 4--5.
In each of these four cases, the proof  fails (for the precondition of \lstinline{g} on line 41 
and/or the assertion on line 39) for different reasons.
In fact, only in the first two cases the code and specification are not compliant, while
the third failure is due to a too weak subcontract, and the last one comes from a prover incapacity.
To the best of our knowledge, 
none of the existing  techniques allows to automatically
distinguish the three reasons and suggest suitable actions.
This work proposes a complete methodology to provide such assistance.
}



\vspace{-2mm}
\section{\modif{Categories} of Proof Failures and Counter-Examples}
\label{sec:definitions}
\vspace{-2mm}


Let $P$ be a C program annotated in \eacsl, 
and $f$ the function under verification in $P$.
Function $f$ is assumed to be recursion-free.
It may call other functions, let $g$ denote any of them.
A \emph{test datum} $V$ for $f$ is a vector of values for all input variables of $f$.
\modif{%
The \emph{program path} activated by a test datum $V$, denoted $\pi_V$,  is the sequence of program statements
executed by the program on the test datum $V$.}
We use the general term of a \emph{contract} 
to designate the set of \eacsl annotations describing a loop or a function.
A function contract is composed of pre- and postconditions including \eacsl clauses
\lstinline{requires}, \lstinline{assigns} and
\lstinline{ensures}  (cf. lines 22--30 in Fig.~\ref{fig:rgf1}).
A loop contract is composed of \lstinline{loop invariant},
\lstinline{loop variant} and \lstinline{loop assigns} clauses
(cf. lines 15--18 in Fig.~\ref{fig:rgf1}).

\modif{Obviously,} an annotation 
cannot be proved for all inputs if there exist inputs
for which the property does not hold.
The notion of \emph{counter-example} depends on the way annotations are evaluated.
The diagnosis of proof failures 
based on the prover's counter-examples can be imprecise
since from the prover's point of view,
the code of callees and loops in $f$ is replaced by the corresponding
subcontracts.
To make this diagnosis more precise, we propose to 
take into account their code as well as their contracts, and to treat both by testing. 
In this section, we define three kinds of 
proof failure reasons, two kinds of counter-examples
and associated detection techniques.
\modif{%
Sec. \ref{subsec:NC} defines a non-compliance 
and briefly recalls the detection technique previously published in~\cite{Petiot/SCAM14}.
Sec. \ref{subsec:SW} is part of the original contribution of this paper,
which introduces too new categories of proof failures and a new translation for test generation.
}

\modif{
\subsection{Non-Compliance}
\label{subsec:NC}
}

A previous work~\cite{Petiot/SCAM14} formally described 
how to transform a C program $P$ annotated in \eacsl
into an instrumented program, denoted  $P^\NC$ in this paper,
on which we can apply test generation to produce test data 
violating some annotations at runtime.\footnote{
This translation 
is illustrated by Fig.\ref{fig:NC-transf} in Appendix~\ref{sec:appendix}.
\modif{For simplicity, we present it for all annotations at the same time as in~\cite{Petiot/SCAM14}.
Its adaptation for a modular approach, or even to a particular annotation whose proof fails,
is straightforward.}}
$P^\NC$ checks \modif{all annotations of $P$}
in the corresponding program locations
and reports any failure.
For instance, the postcondition $\textit{Post}_f$ of $f$
is evaluated by the following code inserted at the end of the function $f$ in $P^\NC$:
\begin{equation}\tag{$\dagger$}
\mbox{\lstinline[mathescape]'int post_f;  $\ Spec2Code$($\textit{Post}_f$, post_f); fassert(post_f);'}
\end{equation}
For an \eacsl predicate \lstinline[mathescape]'$\textit{P}$',
we denote by \lstinline[mathescape]'$Spec2Code$($\textit{P}$, b)'
the generated C code  evaluating the predicate
\lstinline[mathescape]'$\textit{P}$'
and assigning its validity status to the Boolean 
variable \lstinline[mathescape]'b' (see~\cite{Petiot/SCAM14} for details).
The function call \lstinline[mathescape]'fassert(b)'
is expanded into a conditional
statement \lstinline[mathescape]'if(b)' that reports the failure and exits 
whenever \lstinline[mathescape]'b' is false.
Similarly, preconditions and postconditions of a callee $g$
are evaluated respectively before and after executing the function $g$.
A loop invariant is checked before the loop (for being initially true) 
and after each loop iteration (for being preserved by the previous
loop iteration). 
An assertion is checked at its location.
To generate only test data  that respect
the precondition $\textit{Pre}_f$ of $f$,
it is checked  in the beginning of $f$ 
similarly to ($\dagger$)
except that \fassert
is replaced by \fassume
to assume the given condition.

\vspace{-2mm}
\begin{definition}[Non-compliance] 
\label{def:NC}
We say that there is a \emph{non-compliance} between code and specification in $P$
if there exists  a test datum $V$ for $f$ respecting its precondition,
such that $P^{\NC}$ reports an annotation failure on $V$.
In this case, we say that $V$ is a \emph{non-compliance counter-example} (\NCCE).
\end{definition}
\vspace{-2mm}

Test generation on the translated program $P^{\NC}$ 
can be used to generate \NCCE{}s (cf.~\cite{Petiot/SCAM14}).
We call this technique \emph{Non-Compliance Detection} (\NCD).
In this work we use the \pathcrawler test generator that
will try to cover all program paths.
Since the translation step has added a branch for the false
value of each annotation, \pathcrawler will try to cover at
least one path where the annotation does not hold.
(An optimization in \pathcrawler avoids covering the 
same \fassert failure several times.)
The \NCD step may have three outcomes. 
\modif{%
It returns (\nc,\,$V$,\,$a$) if an \NCCE $V$ has been found 
indicating the failing annotation $a$ and recording
the program path $\pi_V$ activated by $V$ on $P^{\NC}$.}
\modif{Second, if it has managed to perform a complete exploration 
of all program paths without finding an \NCCE, it returns \no
(cf. the discussion of completeness in the end of Sec. \ref{sec:framac}).}
Otherwise, if only a partial exploration of program paths
has been performed (due to a timeout,
partial coverage criterion or any other limitation), it returns 
\textsf{?} (unknown).


\begin{figure}[tb]
\begin{minipage}{0.4\columnwidth}
\begin{lstlisting}[mathescape]
/*@ assigns k1,...,kN;
  @ ensures P; */
$Type_g$ g(...){ code1; }




$Type_f$ f(...){ code2;
  g($Args_g$);
  code3; }
\end{lstlisting}
\end{minipage}
\hspace{-6mm}
\begin{minipage}{0.07\columnwidth}$\to$\end{minipage}
\begin{minipage}{0.6\columnwidth}
\begin{lstlisting}[mathescape]
$Type_g$ g_swd(...){
  k1=Nondet(); ... kN=Nondet();
  $Type_g$ ret = Nondet();
  int post; $Spec2Code$(P, post);
  fassume(post); return ret;
} //respects contract of g
$Type_g$ g(...){ code1; }
$Type_f$ f(...){ code2;
  g_swd($Args_g$);
  code3; }
\end{lstlisting}
\end{minipage}
\vspace{-5mm}
\caption{(a) A contract $c\in\C$ of callee $g$ in $f$, vs.
(b) its translation 
for \CWD}
\vspace{-3mm}
\label{fig:CW-transf-functions}
\end{figure}

\begin{figure*}[tb]
\begin{center}
\begin{minipage}{0.8\columnwidth}
\begin{lstlisting}[mathescape]
$Type_f$ f(...){ code1;
  /*@ loop assigns x1,...,xN;
    @ loop invariant I; */
  while(b){ code2; }
  code3; }
\end{lstlisting}
\end{minipage}
\hspace{-6mm}
\begin{minipage}{0.07\columnwidth}$\to$\end{minipage}
\begin{minipage}{0.8\columnwidth}
\begin{lstlisting}[mathescape]
$Type_f$ f(...){ code1;
  x1=Nondet(); ... xN=Nondet();
  int inv1; $Spec2Code$(I, inv1);
  fassume(inv1 &$ $& !b); //respects l$ $oop contract
  code3; }
\end{lstlisting}
\end{minipage}
\vspace{-3mm}
\caption{(a) A contract $c\in\C$ of a loop in $f$, vs. 
(b) its translation 
for \CWD}
\vspace{-3mm}
\label{fig:CW-transf-loops}
\end{center}
\end{figure*}

\modif{
\subsection{Subcontract Weakness and Prover Incapacity}
\label{subsec:SW}
}
\modif{To introduce the new categories of proof failures, we follow
the modular verification approach and need a few definitions.}
%
%
A \emph{non-imbricated} loop (resp. function, assertion) in $f$ is a loop
 (resp. function called, assertion) in $f$ outside any loop in
$f$. A \emph{subcontract for $f$} is the contract of some non-imbricated loop or function
 in $f$. A \emph{non-imbricated annotation} in $f$ is either a
non-imbricated assertion or an annotation in a subcontract for $f$.
For instance, the function $f$ of Fig.~\ref{fig:rgf1} has two subcontracts:
the contract of the called function $g$ and the contract of the loop on lines 33--37.
The contract of the loop in $g$ on lines 15--19 is not a subcontract for $f$, but is a subcontract for $g$.

\modif{We focus on non-imbricated annotations in $f$ and assume}  that all subcontracts for $f$ are respected:
the called functions in $f$ respect their contracts, and the loops in $f$  preserve their loop invariants
and respect all imbricated annotations. 
Let $c_f$ denote the contract of $f$,
$\C$ the set of non-imbricated subcontracts for $f$,
and $\A$ the set of all non-imbricated annotations in $f$ \modif{and the annotations of $c_f$.
In other words, $\A$ contains  
the annotations included in the contracts $\C\cup\{c_f\}$ as well as non-imbricated  assertions in $f$.}
\modif{We also assume that any subcontract of $f$ contains a (loop) assigns clause.
This assumption is not restrictive since such a clause is anyway necessary 
for the proof of any nontrivial code.}

\textbf{Subcontract weakness.}
\modif{To apply testing for}  the contracts of called functions and loops in $\C$
instead of their code,
we use a program transformation of $P$ producing a new program $P^{\GSW}$.
\modif{The code of all non-imbricated function calls and loops in $f$
is replaced by a new one as follows.}

For the contract $c\in\C$ of a called function $g$ in $f$, 
the program transformation  (illustrated by Fig.~\ref{fig:CW-transf-functions})
generates a new function \lstinline{g_swd} with the same signature 
whose code simulates any possible behavior respecting the postcondition in $c$, 
and replaces all calls to $g$ by  a call to \lstinline{g_swd}.
First, \lstinline{g_swd} allows any of the variables
(or, more generally, left-values) present in
the \lstinline{assigns} clause of $c$ to change its value 
(line 2 in Fig.\ref{fig:CW-transf-functions}(b)).
It can be realized by assigning a non-deterministic 
value of the appropriate type
using a dedicated function, denoted here by \lstinline{Nondet()}
(or simply by adding an array of fresh input variables and reading
a different value for each use and each function invocation).
If the return type of $g$ is not \lstinline{void},
another non-deterministic value is read for the returned value
\lstinline{ret} (line 3 in Fig.\ref{fig:CW-transf-functions}(b)).
Finally, the validity of the postcondition is evaluated (taking into
account these new non-deterministic values) 
and assumed
in order to consider only executions that respect 
the postcondition, and
the function returns (lines 4--5 in Fig.\ref{fig:CW-transf-functions}(b)).

Similarly, for the contract $c\in\C$ of a loop in $f$, 
the program transformation  
replaces the code of the loop by 
another code that simulates any possible behavior respecting $c$, that is, 
ensuring the ``loop postcondition'' $I\wedge \neg b$ after the loop as shown in Fig.~\ref{fig:CW-transf-loops}.
In addition, the transformation treats in the same way as in $P^{\NC}$
all other annotations in $\A$:
preconditions of called functions, initial loop invariant verifications 
and the  pre- and postcondition of $f$
(they are not shown 
in Fig. \ref{fig:CW-transf-functions}(b) and \ref{fig:CW-transf-loops}(b)).

\begin{figure}[tb]
  \begin{subfigure}{0.5\textwidth}
\begin{lstlisting}
int x;
/*@ ensures x >= \old(x)+1; assigns x; */
void g1() { x=x+2; }
/*@ ensures x >= \old(x)+1; assigns x; */
void g2() { x=x+2; }
/*@ ensures x >= \old(x)+1; assigns x; */
void g3() { x=x+2; }
/*@ ensures x >= \old(x)+4; assigns x; */
void f() { g1(); g2(); g3(); }
\end{lstlisting}
    \vspace{-2mm}
    \caption{Absence of \CWCE{}s for any single subcontract does not imply absence of global \CWCE{}s}
    \label{fig:contracts-ex1}
  \end{subfigure}
  \hspace{4mm}
  \begin{subfigure}{0.5\textwidth}
\begin{lstlisting}
int x;
/*@ ensures x >= \old(x)+1; assigns x; */
void g1() { x=x+1; }
/*@ ensures x >= \old(x)+1; assigns x; */
void g2() { x=x+1; }
/*@ ensures x >= \old(x)+1; assigns x; */
void g3() { x=x+2; }
/*@ ensures x >= \old(x)+4; assigns x; */
void f() { g1(); g2(); g3(); }
\end{lstlisting}
    \vspace{-2mm}
    \caption{Global \CWCE{}s do not help to find precisely a too weak subcontract}
    \label{fig:contracts-ex2}
  \end{subfigure}
  \vspace{-2mm}
  \caption{Two examples with several subcontracts}
  \vspace{-5mm}
  \label{fig:many-contracts}
\end{figure}

\begin{definition}[Global subcontract weakness] 
\label{def:GSW}
We say that $P$ has a \emph{global subcontract weakness} for $f$ 
if there exists a test datum $V$ for $f$ respecting its precondition,
such that $P^{\NC}$ does not report any annotation failure on $V$,
while $P^{\GSW}$ reports an annotation failure on $V$.
In this case, we say that $V$ is a 
\emph{global subcontract weakness counter-example} (\GSWCE)
for the set  of subcontracts $\C$.
\end{definition}

Notice that we do not consider the same counter-example as an \NCCE  and 
an \SWCE. Indeed, even if some counter-examples may illustrate both a
subcontract weakness and a non-compliance, we consider that
non-compliances usually come
from a direct conflict between the code and the specification
and should be addressed first, while 
contract weaknesses are often more subtle and will be
easier to address when non-compliances are eliminated.

Again, test generation can be applied on  $P^{\GSW}$ to generate
\GSWCE candidates. When it finds a test datum $V$ such that 
$P^{\GSW}$ fails on $V$, we use runtime assertion checking:
if $P^{\NC}$ fails on $V$, then
$V$ is classified as an \NCCE, otherwise  $V$ is a \GSWCE.
We call this technique \emph{Global Subcontract Weakness Detection} for the set of 
all subcontracts, denoted $\GSWD$. 
The $\GSWD$ step may have four outcomes. 
\modif{%
It returns (\nc,\,$V$,\,$a$) if an \NCCE $V$ has been found for the failing
annotation $a$,
and  (\cw,\,$V$,\,$a$,\,$C$) if $V$ has been finally classified as an \SWCE
indicating the failing annotation $a$ and the set of subcontracts $C$.
The program path $\pi_V$ activated by $V$ and leading to the failure 
(on $P^{\NC}$ or $P^{\GSW}$) is recorded as well.}
If the $\GSWD$ has managed to perform a complete exploration 
of all program paths without finding an \GSWCE, it returns \no.
Otherwise, if only a partial exploration of program paths
has been performed it returns \textsf{?} (unknown).


A \GSWCE indicates a global subcontract weakness but
does not explicitly identify which single subcontract $c\in\C$ is too 
weak.
To do that, we propose another program transformation of $P$ into
an instrumented program  $P_c^{\SSW}$.
It is realized by replacing only one non-imbricated function call or loop 
by the code respecting the postcondition of corresponding subcontract $c$ 
(as indicated in Fig. \ref{fig:CW-transf-functions} and \ref{fig:CW-transf-loops})
and transforming other annotations in $\A$ as in $P^{\NC}$.

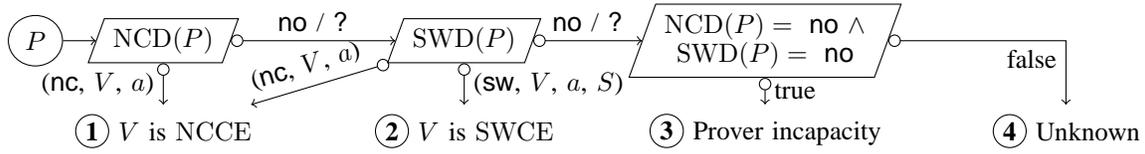
\begin{figure*}[bt]\centering
\begin{tikzpicture}
  \node(p) [data] {$P$};
  \node(ncd) [right of=p,test,node distance=1.7cm] {$\NCD(P)$};
  \node(ncce1) [below of=ncd,node distance=1.2cm] {\circled{1} $V$ is \NCCE};
  \path[darrow] (ncd) -- node[left] {(\nc, $V$, $a$)} (ncce1);
  \path[arrow] (p) -- (ncd);
  \node(cwd) [right of=ncd,test,node distance=4cm] {$\CWD(P)$};
  \path[darrow] (ncd.east) -- node[above] {\no{} / \textsf{?}} (cwd);

  \node(cov-2) [right of=cwd,test,node distance=4cm]{
    $\NCD(P)=~$\no$\land$\\ $\CWD(P)=~$\no};
  \path[darrow] (cwd.east) -- node[above] {\no{} / \textsf{?}} (cov-2);
  \node(pw) [below of=cov-2,node distance=1.2cm]{\circled{3} Prover incapacity};
  \path[darrow] (cov-2) -- node[right] {true} (pw);
  \node(?) [right of=pw,node distance=4cm] {\circled{4} Unknown};
  \path[darrow] (cov-2) -| node[below left] {false} (?);

  \node(cwce-ncce) [below of=cwd,node distance=1.2cm]{\circled{2} $V$ is \CWCE};
  \path[darrow] (cwd) -- node[right] {(\cw, $V$, $a$, $S$)} (cwce-ncce);
  \path[darrow] (cwd) -- node[above,sloped,xshift=-3mm] {~~~(\nc, $V$, $a$)} (ncce1);
\end{tikzpicture}
\vspace{-2mm}
\caption{Combined verification methodology in case of a proof failure on $P$}
\vspace{-5mm}
\label{fig:method-short}
\end{figure*}

\begin{definition}[Single subcontract weakness] 
\label{def:SSW}
Let $c$ be a subcontract for $f$.
We say that $c$ is a \emph{too weak subcontract}
(or has a \emph{single subcontract weakness}) for $f$ 
if there exists a test datum $V$ for $f$ respecting its precondition,
such that $P^{\NC}$ does not report any annotation failure on $V$,
while $P_c^{\SSW}$ reports an annotation failure on $V$.
In this case, we say that $V$ is a \emph{single subcontract weakness counter-example} (\SSWCE)
for the subcontract  $c$ in $f$.
\end{definition}

For any subcontract $c\in\C$, test generation can be separately applied on  $P_{c}^{\SSW}$ to generate
\SSWCE candidates. If such a test datum $V$ is generated, \modif{it is checked on $P^{\NC}$} to 
classify it as an \NCCE or an \SSWCE.
We call this technique,
applied for all subcontracts one after another until a first counter-example  $V$ is found,  
\emph{Single Contract Weakness Detection,}
denoted $\SSWD$. 
The $\SSWD$ step may have three outcomes. 
\modif{%
It returns (\nc,\,$V$,\,$a$) if an \NCCE $V$ has been found for a failing annotation $a$,
and  (\cw,\,$V$,\,$a$,\,$\{c\}$) if $V$ has been finally classified as an \SSWCE
indicating the failing annotation $a$ and the single too weak subcontract $c$.
The program path $\pi_V$ activated by $V$ and leading to the failure 
(on $P^{\NC}$ or $P_c^{\SSW}$) is recorded as well.}
Otherwise, it returns \textsf{?} (unknown), since even after an exhaustive path testing
the absence of \SSWCE for any individual subcontract $c$ does not imply
the absence of a \GSWCE.

Indeed, sometimes $\SSWD$ cannot exhibit a subcontract weakness for a single subcontract
while there is a global subcontract weakness for all of them at once.
For example in Fig.~\ref{fig:contracts-ex1},
if we apply $\SSWD$ to any of the subcontracts, we always have
\lstinline'x >=  \old(x)+5' at the end of $f$
(we add $1$ to $x$ by executing the translated subcontract,
and add $2$ twice by executing the other two functions' code),
so the postcondition of \lstinline'f' holds and no weakness is detected.
If we run  $\GSWD$  to consider all subcontracts at once, we only get
\lstinline'x>= \old(x)+3' after executing the three subcontracts, and can exhibit a counter-example.

On the other hand, running  $\GSWD$
produces a \GSWCE that does not indicate which one of the subcontracts is too weak,
while $\SSWD$ can sometimes be more precise.
For Fig.~\ref{fig:contracts-ex2},
since the three callees are replaced by their subcontracts for  $\GSWD$,  it is impossible
to find out which one is too weak.
\modif{Counter-examples generated by a prover suffer from the same precision issue:
taking into account all subcontracts instead of the corresponding code 
prevents from a precise identification of a single too week subcontract.}
In this example we can be more precise with $\SSWD$,
since only the replacement of the  subcontract of  \lstinline'g3' also leads to an \SSWCE:
we can have \lstinline'x >= \old(x)+3' by executing
\lstinline'g1',  \lstinline'g2' and the subcontract of \lstinline'g3',
exhibiting the contract weakness of \lstinline'g3'.
\modif{Thus, the proposed $\SSWD$  technique can provide the verification engineer
with a more precise diagnostic than counter-examples extracted from a prover.}

We define a combined subcontract weakness detection technique, denoted \SWD, applying
first $\SSWD$ followed by  $\GSWD$ until the first \SWCE is found. 
$\SWD$ may have the same four outcomes as  $\SSWD$.
It allows us 
to be both precise (and indicate when possible a single subcontract being too weak),
and complete (capable to find \GSWCE{}s even when there are no single subcontract weaknesses).

\textbf{Prover incapacity.}
When neither a non-compliance nor a global subcontract weakness
exist, we cannot demonstrate that it is impossible to prove the property.

\begin{definition}[Prover incapacity] 
\label{def:prov-incap}
We say that a proof failure in $P$ is due to a \emph{prover incapacity} 
if for any test datum $V$ for $f$ respecting its precondition,
neither $P^{NC}$ nor $P^{\GSW}$ report any annotation failure on $V$.
In other words, there is no \NCCE and no \GSWCE for $P$.
\end{definition}


\vspace{-2mm}
\section{Diagnosis of Proof Failures using Structural Testing}
\label{sec:global-method}
\vspace{-2mm}

In this section, we present an overview of our method for 
diagnosis of proof failures \modif{using the detection techniques of Sec.~\ref{sec:definitions},
and illustrate it on several examples.
We also provide a comprehensive list of  suggestions of actions for 
each category of proof failures.}

\modif{%
\textbf{The method.}}
The proposed method is illustrated by Fig.~\ref{fig:method-short}.
Suppose that the proof of the annotated program $P$ fails for some non-imbricated annotation $a\in\A$.
The first step tries to find a non-compliance using \NCD. 
If such a non-compliance is found, it generates an \NCCE (marked by \circled{1}
in Fig.~\ref{fig:method-short})
and classifies the proof failure as a non-compliance.
If the first step cannot generate a counter-example,
the \SWD step combines $\SSWD$ and $\GSWD$ 
and tries to generate single \SWCE{}s, then global \SWCE{}s, 
until the first counter-example is generated and classified 
(either as an \NCCE \circled{1} or an \SWCE \circled{2}).
If no counter-example has been found, the last step checks the outcomes.
If both \NCD and \SWD have returned \textsf{no}, that is, 
both  $\NCD$  and $\GSWD$ have performed a complete path exploration 
without finding a counter-example,
the proof failure is classified as a prover incapacity \circled{3} (cf. Def. \ref{def:prov-incap}).
Otherwise, it  remains unclassified \circled{4}.
Fig.~\ref{tab:versions-rgf} associates a variant of the illustrating example to
each case.
\modif{%
For each case, we detail the lines we modified in the program of
Fig.~\ref{fig:rgf1} to obtain a new program, the intermediate results of
deductive verification, \NCD and \SWD and the final verdict (including the
generated counter-example if any).
}

\begin{figure*}[bt]\centering
  \begin{tabular}{c|c|l|c|c|c|c}
    \multirow{2}{*}{\#} & \multicolumn{2}{c|}{Impacted lines}
    & \multicolumn{3}{c|}{Intermediate outcome}
    & \multirow{2}{*}{Final outcome} \\ \cline{2-6}
    & Line & Changes & Proof (failing annot.) & \NCD & \CWD & \\ \hline
    0 & -- & -- & \ok & -- & -- & Proved \\ \hline
    1 & 24 & (deleted) & \textsf{?} (l.39, 41, 26) & \nc & --
    & $V$ = \ce{$\mbox{\lstinline'n=1;' \textbf{\lstinline'a[0]=-214739'}}$}
    is \NCCE \\ \hline

    2 & 34 & \lstinline[style=c]'loop assigns i,a[1..n-1];'
    & \multirow{3}{*}{\textsf{?} (l.39, 41, 42, 26--30)}
    & \multirow{3}{*}{\textsf{?}} & \multirow{3}{*}{\cw{} for l.33--34}
    & $V$ = {\scriptsize{$\langle$}}
    $\mbox{\lstinline'n=2;'} \textbf{\mbox{\lstinline'a[0]=0'}}
    \mbox{\lstinline';a[1]=0;'}$ \\

    &&&&& &$\mathbf{nondet}_{\textbf{\mbox{\lstinline'a[1]'}}}
    \textbf{\mbox{\lstinline'=97157'}}\mbox{\lstinline';'}$ \\
    &&&&& &$\mathrm{nondet}_{\mbox{\lstinline'i'}}\mbox{\lstinline'=0'}$
    {\scriptsize{$\rangle$}} is \SWCE \\ \hline

    3 & 4--5 & (deleted) & \multirow{2}{*}{\textsf{?} (l.39)}
    & \multirow{2}{*}{\no} & \multirow{2}{*}{\no}
    & \multirow{2}{*}{Prover incapacity} \\
    & 22 & \lstinline[style=c]'requires n>0 && n<21;' &&&& \\ \hline
    4 & 4--5 & (deleted) & \textsf{?} (l.39) & \textsf{?} & \textsf{?}
    & Unknown \\
  \end{tabular}
  \vspace{-2mm}
  \caption{Method results for different versions of the illustrating example.}
  \vspace{-5mm}
  \label{tab:versions-rgf}
\end{figure*}

\modif{The proof failure category and the counter-example $V$, along with  
the recorded path $\pi_V$,
the reported failing annotation $a$ and set of too weak subcontracts $S$,
can be extremely helpful for the verification engineer.}
Suppose we try to prove in \Wp a modified version of the function $f$ of
Fig.~\ref{fig:rgf1}
where the precondition at line 24 is missing.
The proof of the precondition of $g$ on line 10 for the call on line 41
fails without indicating a precise reason.
The \NCD step of \stady  generates an \NCCE (case \circled{1},
\#1 in Fig.~\ref{tab:versions-rgf}) where \lstinline'is_rgf(a,n)'
is clearly false due to \lstinline'a[0]' being non-zero, and indicates the
failing annotation (coming from line 10).  
That helps the verification engineer to understand and fix the issue.

Let us suppose now that the clause on line 34 has been erroneously
written as follows: \lstinline'loop assigns i, a[1..n-1];'.
The loop on lines 36--37 still preserves its invariant.
\modif{%
The \NCD step does not find any \NCCE, as this modification did not introduce
any non-compliance between the code and its specification. 
}
Thanks to the replacement shown in Fig.~\ref{fig:CW-transf-loops},
$\SSWD$ for the contract of this loop will detect a single
subcontract weakness for the loop contract (case \circled{2},
\#2 in Fig.~\ref{tab:versions-rgf}),
and report a fail to establish the 
precondition of $g$ (on line 10) for the call on line 41.
With the indication of the single subcontract weakness for the loop, 
the verification engineer will try to strengthen the loop contract
and find the issue.

Suppose now we want to prove the absence of overflow at line 40
of Fig.~\ref{fig:rgf1}, but the lemma on lines 4--5 
(that allows the prover to deduce this property) is missing.
The proof fails  without giving a precise reason since
the prover does not perform the induction needed to deduce the right bounds on
\lstinline'a[i]'.
Neither \NCD nor \CWD can produce a counter-example, and
as the initial program has too many paths, their outcomes are \textsf{?}
(unknown) (case \circled{4}, \#4 in Fig.~\ref{tab:versions-rgf}).
For such situations, \stady offers the possibility to reduce the input domain.
The verification engineer can add the \acsl clause 
\lstinline'typically n<5;' to reduce the array size 
for testing (this clause is ignored by the proof). 
Running \stady now allows the tool to complete the exploration of all
program paths (for \lstinline'n<5') both for \NCD and \CWD without finding a counter-example.
\stady classifies the proof failure for the program with 
the reduced domain as a prover incapacity (case \circled{3},
\#3 in Fig.~\ref{tab:versions-rgf}).
That gives the verification engineer more confidence that the proof failure
has the same reason on the initial program for bigger sizes \lstinline{n}.

The verification engineer prefers to  try interactive proof or adding additional
lemmas or assertions,
and does not waste time looking for a bug or a too week subcontract.

\modif{
\textbf{Suggestions of actions.}
From the possible outcomes of the method illustrated in
Fig.~\ref{fig:method-short} we are able to suggest to the verification engineer 
the most suitable actions (displayed in Fig.~\ref{fig:suggestions})
to help her with the verification task.
A \emph{non-compliance} of the code w.r.t. annotation $a$  means that 
there is an inconsistency between the precondition, the annotation $a$ and the code 
of the path $\pi_V$  leading to $a$.
Thanks to the counter-example, 
the values of variables at different program points along $\pi_V$ 
can be either traced or explored in a debugger \cite{Muller/FM11}. 
In \framac, the execution on $V$ can be 
conveniently explored using \Value or \pathcrawler.
This helps the verification engineer to understand the issue.
Indeed, if an \NCCE is generated, there is no need to
try automatic proof or look for a too weak subcontract --- it will not help.
The reason of the proof failure is necessarily related 
to a non-compliance between 
the code and annotations
traversed by the path $\pi_V$.

A \emph{weakness} of a set of subcontracts $S$ means that at least one of the contracts of $C$
has to be strengthened. By Definitions \ref{def:GSW} and \ref{def:SSW}, the non-compliance is excluded here, 
that is, the execution of $P^{\NC}$ on  $V$  respects the annotation $a$, thus
the suggested action is to strengthen the subcontract(s).
In the case of single subcontract weakness, $S$ is a singleton so the suggestion
is very precise and helpful to the user.
Again, trying interactive proof or additional assertions or lemmas 
will be useless here since the property can obviously not be proved 
because of the counter-example.
For a \emph{prover incapacity,} the verification engineer
may write lemmas or assertions, add hypotheses that may help the theorem prover to
succeed or try another theorem prover.
She also may want to use a proof assistant like \textsc{Coq}, so that she does
not suffer from the limitations of the theorem provers, but 
this task can be more complex and time-consuming.
Finally, when the verdict is \emph{unknown,} test generation for \NCD and/or \SWD times out, 
so the verification engineer may strengthen the
precondition for testing to reduce the input domain, or extend the timeout to
give \stady more time to conclude.
}

\begin{figure*}[bt]\centering
  \modif{
  \begin{tabular}{p{.7cm}|>{\centering\arraybackslash}p{5.8cm}|>{\centering\arraybackslash}p{8cm}}
    \textbf{Case} & {\centering\textbf{Verdict}} & \textbf{Suggestions} \\
    \hline
    \circled{1} & Non-compliance w.r.t. the annotation $a$:
    (\nc, $V$, $a$)
    &
    check the violated annotation $a$
    or the code leading to $a$ in the path $\pi_V$,
    or strengthen the precondition of the function under verification
    \\
    \hline
    \circled{2} & Weakness of subcontracts in $S$ w.r.t. the annotation $a$:
    \ (\cw,~$V$,~$a$,~$S$)
    & strengthen one or several subcontracts in $S$ to exclude the subcontract weakness\\
    \hline
    \circled{3} & Prover incapacity
    & add lemmas or assertions to help the theorem prover,
    or use another prover,
    or an interactive  proof assistant \\
    \hline
    \circled{4} & Unknown
    & strengthen the $\mbox{\lstinline'typically'}$ clause or coverage criterion (e.g. $k$-path),
    or increase the timeout limit for testing \\
  \end{tabular}
  }
  \caption{Suggestions of actions for different categories of proof failures}
  \label{fig:suggestions}
\end{figure*}

\vspace{-2mm}
\section{Implementation and Experiments}
\label{sec:implementation}
\vspace{-2mm}

\begin{figure*}[bt]
  \scriptsize
\mbox{}\hspace{-20mm}
  \begin{center}
  \begin{tabular}{r|c|c|c|c|c|c|c|c|c|c|c|c|c|c|c}
    &&\multicolumn{3}{c|}{Proof}&\multicolumn{4}{c|}{\NCD}
    &\multicolumn{4}{c|}{\CWD}&\multicolumn{2}{c|}{$\NCD+\CWD$}&\\
    \hline
    \input{full_exp_latex_IEEE.csv}
  \end{tabular}
\end{center}
  \caption{Detailed experiments of proof failure diagnosis for mutants with \stady}
  \vspace{-.5cm}
  \label{tab:exp}
\end{figure*}

\textbf{Implementation.}
The proposed method for diagnosis of proof failures 
has been implemented as a \framac plugin, named \stady.
It relies on other plugins: \Wp~\citeframac for deductive
verification and \pathcrawler~\citepathcrawler for structural test generation.
\stady currently supports a significant subset of the \eacsl specification
language, including 
\lstinline'requires', \lstinline'ensures', 
\lstinline'behavior', \lstinline'assumes', 
\lstinline'loop invariant', \lstinline'loop variant' and
\lstinline'assert' clauses.
Quantified predicates
\lstinline[style=c]'\exists' and \lstinline[style=c]'\forall' and builtin terms
as \lstinline'\sum' or \lstinline'\numof' are translated as loops. 
Logic functions and named predicates are treated by inlining.
The \lstinline'\old' constructs are treated by saving the initial
values of formal parameters and global variables at the beginning of the
function. 
Validity checks of pointers are
partially supported due to the current limitation of the underlying test
generator: we can only check the validity of input pointers and global arrays.
The \lstinline'assigns' clauses are only taken into consideration during the
\CWD phase: we do not aim to find what is missing in the \lstinline'assigns'
clause (\NCD) because provers usually give sufficiently good feedback about it,
but we want to find what is unnecessary and could be removed from an
\lstinline'assigns' clause (\CWD).
Inductive predicates, recursive functions and floating-point numbers are
currently not supported and are part of our future work.

\textbf{The research questions} we address in our experiments are the following.

\vspace{-2mm}
\begin{itemize}
\item[\textbf{RQ1}]
Is \stady able to precisely diagnose most proof failures in C programs?
\item[\textbf{RQ2}]
What are the benefits of the \CWD extension (in particular, with respect to
\NCD)?
\item[\textbf{RQ3}]
Is \stady able to generate  \NCCE{}s or \CWCE{}s even with a partial testing coverage?
\item[\textbf{RQ4}]
Is \stady's execution time comparable to the time of an automatic proof?
\end{itemize}
\vspace{-2mm}

\textbf{Experimental protocol.} 
The evaluation used 20 annotated programs from \cite{ACSLbyExample},
whose size varies from 35 to 100 lines of annotated C code.
These programs manipulate arrays, they are fully specified in \acsl and their
specification expresses non-trivial properties of C arrays. 
To evaluate the method
presented in Sec.~\ref{sec:global-method} and its implementation, we apply \stady on systematically
generated altered  versions (or \emph{mutants}) of correct C programs.
Each mutant program is obtained by performing a single modification (or \emph{mutation}) on the
initial program.
The mutations include: a binary operator modification in the code or in the
specification, a condition negation in the code, a relation modification in the
specification, a predicate negation in the specification, a partial loop invariant or
postcondition deletion in the specification.
In this study, we do not mutate the precondition of the function under verification, 
and restrict possible mutations on binary operators to avoid creating absurd
expressions, in particular for pointer arithmetics.

The first step tries to prove each mutant using \Wp. 
The proved mutants respect the specification and are classified as correct. 
Second, we apply the \NCD method on the remaining mutants.
It classifies proof failures for some mutants as non-compliances, indicates the failing annotation and an \NCCE.
The third step applies the \CWD method on remaining mutants,
classifies some of them as subcontract weaknesses, indicates the weak subcontract and a \CWCE.
If no counter-example has been found by the \CWD, the mutant remains 
unclassified. 
The results are displayed in Fig.~\ref{tab:exp}.
The columns 
present the number of generated mutants, and the results of each of the three
steps: the number (\#) and ratio (\%) of classified mutants,
maximal and average execution time (put on two lines) of the step
over classified mutants ($t^\text{\ok}$ or $t^\text{\ko}$) and over non-classified
mutants ($t^\text{?}$) at this step.
The ratios are computed with respect to unclassified mutants after the previous step.
The $\NCD+\CWD$ columns sum up selected results after both $\NCD$ and $\CWD$ steps:
the average and maximal time ($t$) are shown globally over all mutants.
The time is computed until the proof is finished or until the first counter-example is generated.
The final number of remaining unclassified mutants (\#?) is given in the last column.

\textbf{Experimental results.}
For the 20 considered programs, 928 mutants have been generated. 80 of them 
have been proved by \Wp.
Among the 848 unproven mutants, \NCD has detected a non-compliance
induced by the mutation in 776 mutants (91.5\%),
leaving 72 unclassified.
Among them, \CWD has been able to exhibit a counter-example (either a \NCCE or a
\CWCE)
for 48 of them (66.7\%), finally leaving 24 programs unclassified.
They can be either equivalent mutants that were not proved
by \Wp due to a prover incapacity, or mutants coming from a mutation 
in an unsupported annotation being undetectable by the
current version, or incorrect mutants for which testing was incomplete due to a timeout.
Regarding \textbf{RQ1}, \stady has found a precise reason
of the proof failures  and produced a counter-example 
in 824 of the 848 unproven mutants, 
i.e. classifying 97.2\%.
Exploring the benefits of detecting a prover incapacity may often require to manually reduce
the input domain, to try additional lemmas or interactive proof, so it   
was not sufficiently investigated in this study 
(and would probably require another, non mutational approach).

Regarding \textbf{RQ2}, 
\NCD alone diagnosed 776 of 848 unproven mutants (91.5\%).
\CWD diagnosed 48 of the 72 remaining  mutants (66.7\%)
bringing a significant complementary contribution 
to a better understanding of reasons of many proof failures.

In our experiments,
each prover can try to prove each verification condition 
during at most 40 seconds.
We also set a timeout for any test generation session
to 5 seconds, i.e. one session for the \NCD step, and 
several sessions for \CWD steps.
We also limit the depth of explored program paths with the 
{\em k-path} criterion (cf. Sec. \ref{sec:framac})
setting $k = 4$.
Both the session timeout and the {\em k-path} heavily limit the testing coverage
but \stady still detects 97.2\% of faults in the generated programs.
That addresses \textbf{RQ3} and demonstrates that the proposed method can efficiently 
classify proof failures and generate counter-examples
even with a partial testing coverage and can therefore 
be used for programs where the 
total number of paths cannot be limited
(e.g. by the \lstinline{typically} clause).

Concerning \textbf{RQ4},
on the considered programs \Wp needs on average 2.6 sec. per mutant (at most 4.4 sec.) to
prove a program, and spends 13.0 sec. on average (at most 61.3 sec.) when the
proof fails.
The total execution time of \stady is comparable: it needs on average 2.7 sec.  per unproven mutant 
(at most 19.9 sec.).

\textbf{Summary.}
The experiments show that the proposed method can automatically classify a significant number
of proof failures within an analysis time comparable to the time of an automatic proof
and for programs for which only a partial testing coverage is possible.
The \CWD technique offers an efficient complement to \NCD for a more 
complete and more precise diagnosis of proof failures.

\modif{
\textbf{Threats to validity.}
As it is often the case in software verification studies, one major threat is
related to the 
representativeness of results, i.e. their \textit{external validity}.
In our case, due to the nature of the problem,
we are restricted to realistic annotated programs
that cannot be generated automatically 
or extracted from existing databases of unspecified code.
Therefore, to reduce this threat, we used programs from an \textit{independent}
benchmark \cite{ACSLbyExample} created in order to illustrate
on different examples the usage of the \acsl specification 
language for deductive verification with \framac.


\textit{Scalability} of the results is another threat
since we do not demonstrate their validity for functions of larger programs. 
Because of the modular reasoning of deductive verification,
it can be argued that the proposed technique should only be applied on a unit level,
separately for each function, since the verification engineer proves a program  in this way.
Indeed, in the current practice of deductive verification, it does not make sense to analyze
proof failures for the whole module or application at the same time.

The main scalability concern is thus related to the usage of structural test generation 
that can often time out without achieving a full coverage.
To address this issue, we have specifically investigated the impact of a partial test coverage
on the effectiveness of the method (cf. \textbf{RQ3} above) and proposed
a convenient way to reduce the input domain (using \lstinline{typically} clause, 
an extension of \acsl).

Other threats can be due to the used measurements, i.e. \textit{construct validity}.
To reduce this threat, we used a careful measurement of 
results (including analysis time for each step and 
each mutant, their mean and maximal values, 
separately computed for classified and unclassified proof failures).
One concern is producing realistic situations in which 
the verification engineer can need help in the analysis of proof failures.
While the first users of \stady have appreciated its feedback, 
we have not yet had the opportunity to organize a fair evaluation with a 
representative group of users. 
Thus we have performed an extended set of experiments using simulation of 
errors by mutations as an alternative in the meanwhile. 
We have chosen a large subset of mutation operators (mutation in the code,
mutation in an annotation, deletion of an annotation) that model 
frequent problematic situations 
(incorrect code or annotations, incomplete specification)
leading to proof failures.
This approach looks suitable for non-compliance and subcontract weaknesses, and
certainly less suitable for the more subtle prover incapacity cases.
The results should be later confirmed by a representative user study.
}

\vspace{-2mm}
\section{Related Work}
\label{sec:related}
\vspace{-1mm}

\textbf{Understanding proof failures.}
A two-step verification in~\cite{Tschannen/14} compares the proof failures of an
Eiffel program with those of its variant  where called functions are
inlined and loops are unrolled. It reports code and contract revision
suggestions from this comparison. Inlining and unrolling are respectively
limited to a given number of nested calls and explicit iterations. If that
number is too small the semantics is lost and a warning of unsoundness is also
reported to the user.

\textbf{Proof tree analysis.}
More precision can be statically obtained by analyzing the unclosed branches of
a proof tree. The work \cite{Gladisch/TAP09} is performed 
in the context of \textsc{KeY} and its verification calculus
that applies  deduction rules to a dynamic formula mixing  a program and its
specification. It proposes \textit{falsifiability preservation
checking} that helps to distinguish whether the branch failure
comes from a programming error or from a contract weakness. However this
technique can detect bugs only if contracts are strong enough. Moreover it is automatic
only if a prover (typically, an SMT solver) can decide the non-satisfiability of
the first-order formula expressing the falsifiability preservation condition.
\cite{Engel/TAP07} exploits the proof trees generated during a
proof attempt by \textsc{KeY}. The relevance of generated tests depends on the quality of the
specification written by the user, 
and it does not allow to distinguish non-compliances from
specification weaknesses.

\textbf{Combination of static and dynamic analysis.}
Static and dynamic analysis work better when used together, as in the method
\textsc{Synergy}~\cite{Gulavani/06}, its interprocedural and compositional
extension in \textsc{Smash}~\cite{Godefroid/POPL10}, the method \sante~\citesante
 and the present method. Static analysis maintains an
over-approximation that aims at verifying the correctness of the system, while
dynamic analysis maintains an under-approximation trying to detect an error.
Both abstractions help each other in a way similar to the counter-example guided
abstraction refinement method (\textsc{Cegar})~\cite{Clarke/03}.



\textbf{Counter-examples for non-inductive invariants.}
Counter-examples can be generated to show that invariants proposed for
transition systems are too strong or too weak~\cite{Claessen/TAP08}.
Differences with our work are the focus on invariants, the formalism of
transition systems, and the use of random testing (with \textsc{QuickCheck}).

\textbf{Other verification feedbacks.}
Our goal was to find input data to illustrate proof failures.
A complementary work \cite{Muller/FM11} proposed to extend a runtime assertion
checker to use it as a debugger to help the user understand complex proof
failure counter-examples.
The \textsc{Dafny} development environment \cite{Leino/FIDE14} provides
verification feedback to the user during the programming phase. It integrates
the \textsc{Boogie} Verification Debugger \cite{LeGoues/SEFM11}
that helps the understanding of verification tools like \textsc{Boogie}.
Currently, \textsc{Dafny} only uses counter-examples provided by the solver, 
and does not produce as much information when verification
times out as it does when verification fails.

\textbf{Checking prover assumptions.} 
Axioms are logic properties 
 used as hypotheses by 
provers and thus usually not checked.
Model-based testing applied to a computational model of an axiom permits to
detect errors in axioms and thus to maintain the soundness of the
axiomatization~\cite{Ahn/TAP10}. 
This work is complementary to ours because it tackles the case of
deductive verification trivially succeeding due to an invalid axiomatization,
whereas we tackle the case of inconclusive deductive verification.
\cite{Christakis/FM12} proposed to complete the results of static checkers with
dynamic symbolic execution using \textsc{Pex}.
The explicit assumptions used by the verifier (absence of overflows,
non-aliasing, etc.) create new branches in the program's control flow graph
which \textsc{Pex} tries to explore.
This approach permits to detect errors out of the scope of the considered static
checkers, but does not provide counter-examples in case of a specification
weakness.

\smallskip
\modif{%
\textbf{The present work} continues the previous efforts to facilitate deductive verification
by generating  counter-examples.
We propose an original detection technique of three  categories of proof failure 
that gives  
a more precise diagnostic than in the previous work using testing.
Thanks to the separate detection of non-compliances and single subcontract weaknesses, 
the generated counter-examples can better identify the reasons of proof failures
than those extracted from a solver.
To the best of our knowledge, such a complete testing-based  methodology 
proposed in this paper,
automatically providing  to the verification engineer 
a precise feedback on proof failures 
was not studied, implemented and evaluated before.}

\vspace{-2mm}
\section{Conclusion and Future Work}
\label{sec:conclusion}
\vspace{-2mm}

We proposed a new approach to improve the user feedback in case of a proof failure.
Our method relies on test generation and helps to
decide whether the proof has failed or timed out due to a non-compliance between
the code and the specification, a subcontract weakness, or a prover weakness.
\modif{%
This approach is based on a spec-to-code program transformation that allows to
use a test generator taking a C program as input.
The transformation 
for \SWD is an original contribution of this paper.
}
Our experiments show that our implementation -- as a \framac plugin, \stady --
was able to diagnose over 97\% of the programs (generated by introducing a
mutation in a verified program).

One benefit of the proposed approach is the capacity to provide the verification engineer 
with a precise reason of a proof failure that helps to choose the right way to proceed
and facilitates the processing of proof failures.
Counter-examples illustrate the issue on concrete values and 
help to find out more easily why the proof fails.
The method is completely automatic, relies on the existing specification  
and does not require any additional manual specification or instrumentation task.
As a consequence, this method can be adopted by less experienced verification 
engineers and software developers.

One requirement of the complete method coming from test generation is to have the C code of 
called functions, while the $\GSWD$ technique  remains applicable even without 
source code.
Another limitation is related to a potentially very big number of program path, that
cannot be explored. 
Initial experiments show that proof failures can be classified in practice
even after test generation with a partial test coverage, within a testing time
comparable to the time of the proof.

We are convinced that the proposed methodology facilitates the verification task
by lowering the level of expertise required to conduct a deductive program
proof, removing one of the major obstacles for a wider use of deductive
verification in industry.
Future work includes further evaluation of the proposed methodology,
\modif{a study of optimized combinations of \NCD and \SWD for subsets of annotations and subcontracts,
experiments on a larger class of programs and a better support of \eacsl constructs in
our implementation (inductive predicates, validity of non-input pointers).}

\smallskip
\noindent
\textit{Acknowledgment.}
The authors thank the \framac and \pathcrawler teams for providing the
tools and support.
Special thanks to 
Fran\c{c}ois Bobot, 
Lo\"{i}c Correnson, 
Julien Signoles
and
Nicky Williams 
for many fruitful discussions, suggestions and advice.

\bibliographystyle{IEEEtran}
\bibliography{biblio}

\newpage

\appendix
\vspace{-2mm}
\section{Appendix}
\label{sec:appendix}
\vspace{-1mm}

\textbf{Program transformation for non-com\-pli\-ance detection.}
Fig.~\ref{fig:NC-transf} illustrates the translation of an annotated program $P$
into another C program, $P^{\NC}$, that is used to generate counter-examples 
during non-compliance detection (\NCD).

\begin{figure}[tb]
\begin{minipage}{0.4\columnwidth}
\begin{lstlisting}[mathescape]
/*@ requires P1;
    ensures P2; */
$Type_g$ g(...) {
  code1;
}
/*@ requires P5;
    ensures P6; */
$Type_f$ f(...) {
  code2;
  g(...);
  //@ loop invariant P3;
  while(b) {
    code3;
  }
  code4;
  //@ assert P4;
  code5;
}
\end{lstlisting}
\end{minipage}
\hspace{-6mm}
\begin{minipage}{0.07\columnwidth}$\to$\end{minipage}
\begin{minipage}{0.6\columnwidth}
\begin{lstlisting}[mathescape]
$Type_g$ g(...) {
  int pre_g; $Spec2Code$(P1, pre_g);
  fassert(pre_g);
  code1; 
  int post_g; $Spec2Code$(P2,post_g);
  fassert(post_g);
}
$Type_f$ f(...) {
  int pre_f; $Spec2Code$(P5, pre_f);
  fassume(pre_f);
  code2;
  g(...);
  int inv1; $Spec2Code$(P3, inv1);
  fassert(inv1);
  while(b) {
    code3;
    int inv2; $Spec2Code$(P3, inv2);
    fassert(inv2);
  }
  code4;
  int asrt; $Spec2Code$(P4, asrt);
  fassert(asrt);
  code5;
  int post_f; $Spec2Code$(P6,post_f);
  fassert(post_f);
}
\end{lstlisting}
\end{minipage}
\vspace{-5mm}
\caption{(a) An annotated code, vs.  (b) its translation 
for \NCD}
\vspace{-3mm}
\label{fig:NC-transf}
\end{figure}

\end{document}